\documentclass[prd,amsmath,amssymb,nofootinbib]{revtex4}
\usepackage{amsmath,amssymb}
\usepackage{parskip}
\usepackage{graphicx}
\usepackage[normalem]{ulem}
\def\x{{\boldsymbol x}}

\def\0{{\boldsymbol 0}}
\def\p{{\boldsymbol p}}
\def\q{{\boldsymbol q}}

\def\l<{\left<} \def\r>{\right>}
\def\wt{\widetilde}

\usepackage[dvips]{color} 

\renewcommand\sout{\bgroup \color{red} \ULdepth=-.5ex \ULset}

\begin{document}

\begin{flushright}
KEK-TH-1670\\
UT-Komaba-13-12
\end{flushright}

\vspace*{1cm}

\title{Microscopic identification of dissipative modes 
in relativistic field theories}

\author{Yohei Saito}
\affiliation{KEK Theory Center, IPNS,
          High Energy Accelerator Research Organization (KEK) \\
           1-1 Oho, Tsukuba, Ibaraki, 305-0801, Japan}

\author{Hirotsugu Fujii}
\affiliation{Institute of Physics, University of Tokyo, Komaba 3-8-1, Tokyo 153-8902, Japan}

\author{Kazunori Itakura}
\affiliation{KEK Theory Center, IPNS,
          High Energy Accelerator Research Organization (KEK) \\
           1-1 Oho, Tsukuba, Ibaraki, 305-0801, Japan}
\affiliation{Department of Particle and Nuclear Studies, 
Graduate University for Advanced Studies (SOKENDAI), 
1-1 Oho, Tsukuba, Ibaraki 305-0801, Japan}

\author{Osamu Morimatsu}%
\affiliation{KEK Theory Center, IPNS,
          High Energy Accelerator Research Organization (KEK) \\
           1-1 Oho, Tsukuba, Ibaraki, 305-0801, Japan}
\affiliation{Department of Particle and Nuclear Studies, 
Graduate University for\\ Advanced Studies (SOKENDAI), 
1-1 Oho, Tsukuba, Ibaraki 305-0801, Japan}
\affiliation{Department of Physics, Faculty of Science, University of Tokyo,\\ 7-3-1 
Hongo Bunkyo-ku Tokyo 113-0033, Japan\\}

\begin{abstract}
We present an argument to support the existence of dissipative 
modes in relativistic field theories. In an $O(N)$ $\varphi^4$ 
theory in spatial 
dimension $d\le 3$, a relaxation constant $\Gamma$ of 
a two-point function in an infrared region is shown to be finite 
within the two-particle irreducible (2PI) framework at the 
next-leading order (NLO) of $1/N$ expansion.
This immediately implies that a slow dissipative mode 
with a dispersion $p_0\sim i\Gamma \p^2$ is 
microscopically identified in the two-point function. 
Contrary, NLO calculation in the one-particle irreducible (1PI) 
framework fails to yield a finite relaxation constant. 
Comparing the results in 1PI and 2PI frameworks, one concludes that 
dissipation emerges from multiple scattering of a particle with 
a heat bath, which is appropriately treated 
in the 2PI-NLO calculation through the resummation of secular terms to
improve long-time behavior of the two-point function. Assuming that 
this slow dissipative mode survives 
at the critical point, one can identify the dynamic critical exponent $z$
for the two-point function as $z=2-\eta$. We also discuss possible 
improvement of the result.

\end{abstract}

\maketitle

\newpage

\section{Introduction}


It is generally a non-trivial and difficult challenge to 
deduce from a microscopic theory hydrodynamic behavior
at long-time scales. As was clearly stated by 
Kadanoff and Martin \cite{AnnPhys.24.419}, it is because 
``{\it hydrodynamic 
equations only appear when the behavior is dominated by the secular 
effects of collisions}" and thus ``{\it most straightforward techniques for 
determining the correlation functions cannot be successfully applied}". 
In the present paper, we will demonstrate in a concrete model 
that the two-particle 
irreducible (2PI) framework which appropriately deals with 
secular effects through resummation is indeed able to 
describe the hydrodynamic behavior from the microscopic level, 
in particular, a dissipative mode in a two-point correlation function.

There are three different levels of descriptions for long-time 
(and long-distance) behavior of a locally equilibrated system. 
At time scales much longer than the correlation time (typical time scale 
of the system),
hydrodynamics is established as the framework for  
time evolution of conserved densities through macroscopic variables 
such as temperature, chemical potentials and flow velocities.
Diffusion and viscosity constants are regarded as low-energy effective
constants.
When we study (e.g.) temperature or momentum dependence of 
the low-energy constants,
we need to include fluctuations around hydrodynamic behavior
by employing a ``meso-scopic'' effective theory. 
The Mode-Coupling Theory (MCT)
\cite{PhysRevLett.19.700, PhysRev.177.952, Ann.Phys.61.1, Hohenberg:1977ym}
as a typical example, separates the slow and fast degrees
of freedom and describes effective dynamics of the slow modes with
treating effects of 
the fast modes as stochastic noise and bare parameter constants.
In particular, non-conserved variables such as an order parameter 
field $\varphi$ 
may be included in addition to hydrodynamic modes in MCT 
when they contribute to slow dynamics of interest.
Only 
after the microscopic theory is solved,
all the phenomenological parameters would be fixed,
which is the third level description.
However, such a microscopic approach,
if applied naively, is suitable only for short time description 
but starts to lose its accuracy at time scales longer 
than typical scattering time.

It should be also noticed that hydrodynamics applies at the scale 
much longer than the correlation length $\xi$
and therefore breaks down as the system approaches a critical point,
where $\xi \to \infty$. 
MCT was originally devised to describe the slow dynamics
near the critical point by including the order-parameter fluctuations
$\varphi$ as well as those of conserved densities,
and it is successful to reproduce the singularity of the transport 
coefficients as the critical point is approached.
Accordingly MCT gives empirical values for the dynamic critical exponent $z$, 
which characterizes the dissipation time of a disturbance
as $\tau \sim \xi^z$.
Such a process should be associated with
a dissipative mode whose dispersion is given by $p_0 \propto i |\p|^z$
with $p_0$ and $\p$ the mode energy and momentum. 
The central problem of this paper is
how to identify this dissipative mode 
by analyzing the two-point function of a microscopic theory.

Considering that MCT is just a phenomenological theory
and that the dissipative mode (at the critical point) should couple
in the two-point function of the order-parameter field $\varphi$, 
it is preferable if we can investigate the low-frequency behavior
of the two-point function at the microscopic level. 
As far as we know, 
the 2PI approach \cite{Luttinger:1960ua, Cornwall:1974vz, Berges:2004yj} 
is the only framework which allows us to describe the
two-point function at the microscopic level, and making the 
secular effects from multiple scattering of particles tractable 
through resummation. This motivated us 
to work on the two-point function of $\varphi$ in the 2PI framework.

Recently, the 2PI effective action has been actively applied to 
non-equilibrium phenomena. One of the merits of the 2PI framework 
is that it treats, in addition to a condensate $\phi=\l< \varphi \r>$, 
the two-point correlation function of fluctuations 
$G=\l<\delta\varphi \delta\varphi \r>$ with $\delta\varphi=\varphi-\phi$ 
in a self-consistent way, and that we can systematically resum 
higher-order diagrams for $\phi$ and $G$ keeping conservation laws 
of the system preserved 
\cite{Baym:1961zz, Baym:1962sx, kadanoff1962quantum}. 
Therefore, the 2PI framework is expected to be 
useful in describing time evolution of conserved densities, and thus 
in describing long-time evolution of the system.

In the present paper, we work on a relativistic $O(N)$ $\varphi^4$ 
scalar theory. 
Another motivation for our study 
is to see how a dissipative mode appears in the microscopic two-point
function of a {\it relativistic} field theory
\footnote{
Concerning the mesoscopic description based on MCT, there
are previous studies 
in nonrelativistic 
\cite{Hohenberg:1977ym, PhysRevLett.29.1548, PhysRevLett.32.1289, 
PhysRevB.10.139, PhysRevB.13.4119}
and relativistic 
\cite{Rajagopal:1992qz, PhysRevD.85.096007, Ohnishi:2004eb} 
field theories. 
All the papers except for Ref.~\cite{Ohnishi:2004eb}
assumed the existence of dissipative modes ($p_0\sim -i |\p|^2$) 
and
obtained the result $z\sim 2$ at the leading order 
of $1/N$ or $\epsilon$ expansion. 
Ref.~\cite{Ohnishi:2004eb} obtained a different result 
$z\sim 1$ in a relativistic model. The authors of Ref.~\cite{Ohnishi:2004eb}
claimed that MCT for relativistic models should be different from that for
nonrelativistic models in the point 
that a ``momentum" variable conjugate 
to a field variable should be included in the MCT for relativistic models. 
We will not address this problem since our scope of the present paper is 
to find a microscopic description of dissipative modes.
}
.
There is a controversial situation concerning 
the dynamic critical exponent $z$ between {\it relativistic}
and {\it nonrelativistic} scalar theories. 
Analysis of the two-point function in a nonrelativistic $O(N)$ model 
\cite{PTP.52.1135, PTP.52.1463} predicts $z$ close to 2, while 
a relativistic model gives different values for $z$ depending on 
the calculation methods: Dynamic renormalization group 
\cite{PhysRevD.63.045007, Boyanovsky:2003ui} gives 
$z\sim 1$ while numerical simulation \cite{Berges:2009jz} 
gives $z\sim 2$. 
In fact, as is shown in the present paper, 
these analyses of the two-point function of $\varphi$, 
except for the last numerical one, do not give us the exponent $z$ of 
a purely dissipative mode, but of the 
bare propagating modes, 
which decay much faster than dissipative modes.
The dissipative mode emerges in the two-point function of $\varphi$
non-perturbatively only after the resummation. 
We will clarify these points in the text.

Before closing Introduction, let us list up 
several physical situations whose hydrodynamics are of interest.
First, hydrodynamic picture is widely applied to describing 
long-time and long-distance behavior of various systems from
Quark-Gluon Plasma (QGP) in heavy-ion collisions
\cite{kolb2004hydrodynamic,hirano2010hydrodynamics,hirano2011hydrodynamic} 
and Bose-Einstein condensate of cold atomic gas 
\cite{PhysRevLett.77.2360}, 
to astrophysical phenomena such as 
supernova explosions \cite{takiwaki2012three},
and even to the universe itself
\cite{weinberg1972gravitation}. 
In particular, 
a critical point is conjectured in the QCD phase diagram,
where the transition from hadronic degrees of freedom to QGP
is of the 2nd order,
and heavy-ion collision experiments for the critical point survey
have already started \cite{odyniec2010rhic, mohanty2011star}. 
Locating the critical point on the QCD phase diagram is 
one of the main objectives in QGP physics now,
and to this end we obviously need correct understanding
of the dynamics near the critical point.
There are indeed several works applying MCT to the 
critical dynamics of QCD\cite{Rajagopal:1992qz, PhysRevD.67.094018, 
PhysRevD.70.014016, PhysRevD.70.056001, Ohnishi:2004eb, PhysRevD.85.096007}.
However, 
it is still controversial in the sense which hydrodynamic mode
dominates the critical dynamics. 
Note also that people are trying to describe non-equilibrium 
phenomena from microscopic theories by exploiting the 2PI framework. 
For example, people discuss the problem of how QGP can be 
formed within a short time after heavy-ion collisions 
\cite{berges2009turbulence, berges2013turbulent, PhysRevD.86.085040, epelbaum2013onset} 
and also non-equilibrium phenomena in cold atomic systems 
\cite{gasenzer2009ultracold}. 
We hope that fundamental understanding on the dissipative modes 
in the present paper would be helpful for various phenomena including 
the QGP physics.

This paper is organized as follows: 
In Section 2, we discuss how a dissipative mode emerges 
from a two-point correlation function using both 1PI and 2PI 
effective actions. In particular, we focus on the way how the effects of 
multiple scattering with a heat bath is included, and propose a criterion
for the emergence of a dissipative mode. 
In Section 3, we check the criterion in $O(N)$-symmetric $\varphi^4$ model 
within two different ways of calculations: 1PI and 2PI frameworks.
We will find that the two-point function in 1PI-NLO calculation 
fails to describe the dissipative mode, but that 2PI-NLO calculation 
can have a dissipative mode with a finite relaxation constant. 
In Section 4, assuming that a dissipative two-point correlation function 
obtained in the previous section is not modified at the critical point, 
we estimate the dynamic critical exponent $z$. We also discuss possible
improvement of our result. The last section is devoted to summary and 
discussion.

\section{Dissipative mode in two-point correlation function}

We consider a system slightly disturbed from 
equilibrium  and
how it undergoes relaxation at time scales much longer
than the typical scattering time, 
by inspecting the two-point function. 
\footnote{
When a system is in the vicinity of equilibrium, 
relaxation is described by the retarded two-point function 
evaluated in equilibrium 
(linear response theory) 
\cite{kubo1957statistical, le2000thermal}.
}
Before proceeding to a concrete model analysis,
let us first clarify general properties of 
two-point functions in an infrared region.

The slowest motion of the system is governed by 
the hydrodynamic modes consisting
of conserved density fluctuations as well as the order parameter
fluctuations in the case of critical phenomena and the Nambu-Goldstone
modes in the case of the symmetry-broken phase.
At ``mesoscopic" time scale,
MCT describes relaxation processes phenomenologically 
as an effective theory which involves other slow degrees of freedom
in addition to the hydrodynamic modes.
Non-linear couplings among those modes
renormalize (e.g.,) transport coefficients in hydrodynamics.

In the present paper we deal with 
a non-conserved ``order parameter" field $\varphi$ in the symmetric
phase, for a particular example. According to the phenomenological MCT,
the bare two-point correlation function of the field 
$\varphi$ with a mass $m$ is given as 
\cite{Hohenberg:1977ym, forster1975hydrodynamic, PhysRevB.10.139, 
PhysRevB.13.4119}\footnote{
For a {\it conserved} field, the two-point function is written as
$  C_{\rm pheno}(p_0,\p)=
   1/[-i p_0/(\Gamma\p^2)+\p^2+m^2] .
$ 
}
 \begin{eqnarray}
 G_{\rm pheno}(p_0,\p)=\frac{1}{-ip_0 /\Gamma + \p^2+m^2}\, . 
\nonumber
 \end{eqnarray}
Here $\Gamma$ is the relaxation constant which is easily 
understood after the inverse Laplace transform with respect to $t$: 
$ G_{\rm pheno}(t,\p)\sim {\rm e}^{-\Gamma\cdot (\p^2+m^2)t}\, .$
On the other hand, 
in a microscopic field theory, a retarded two-point function 
$G_{\rm R}(p_0,\p)$ is generally written as
 \begin{eqnarray}
  G_{\rm R}(p_0,\p) = \frac{1}{-p_0^2+\p^2+m_0^2-\Sigma_{\rm R}(p_0,\p)}\, , 
  \label{fromal-G}
 \end{eqnarray}
where $\Sigma_{\rm R}(p_0,\p)$ is a retarded self-energy 
and $m_0$ is a bare mass. 
Henceforth, we indicate retarded and advanced quantities
with indices R and A, respectively. 
Since the real and imaginary parts of $\Sigma_{\rm R}$ are 
respectively even and odd functions of $p_0$, 
one can expand $\Sigma_{\rm R}$  in the infrared region,
presuming ${\cal O}(p_0) = {\cal O}(\p^2)$,
\begin{eqnarray}
  \Re{\rm e}\,\Sigma_{\rm R}(p_0,\p) &=&
\left(c_0 + d_0\p^2 + {\cal O}(\p^4)\right) + {\cal O}(p_0^2)\, , 
\nonumber \\
  \Im{\rm m}\,\Sigma_{\rm R}(p_0,\p) &=&
\left(c_1 + {\cal O}(\p^2)\right) p_0 +{\cal O}(p_0^3)\, . 
  \label{expand_sigma}
\end{eqnarray}
Here it should be understood that the dimensions 
in ${\cal O}(p_0^n)$ and ${\cal O}(\p^n)$ are
canceled with other dimensionful quantities such as
temperature $T$ and/or $m_0$ 
(and coupling constant $\lambda$ in general),
{\it i.e.,} ${\cal O}\left((p_0/T)^n\right)$, ..., etc.
When we refer to an ``infrared region", it is always meant that 
the energy and momentum are much less than these dimensionful parameters.
Note that the coefficients $c_0,\ c_1$ and $d_0$ depend on 
$\lambda$ and $T$.\footnote{Therefore, $1-d_0$ does not vanish identically.
As we will see, coefficients $c_0$ and $(c_1, d_0)$ 
depend on the coupling constant $\lambda$ and 
are of the order of $N^0$ and $N^{-1}$, respectively,
in the $O(N)$ scalar model.} 
With the expression Eq.~(\ref{expand_sigma}) valid in the 
infrared region,
the two-point function of the microscopic theory
can be approximated as
 \begin{eqnarray}
 G_{\rm R}(p_0,\p)\sim
 \frac{1} {-i c_1 p_0+(1-d_0)\p^2+ m_0^2-c_0}\, .
  \label{hydro-prop}
 \end{eqnarray}
From Eq.~(\ref{hydro-prop}) we see that
the two-point function $G_{\rm R}(p_0,\p)$ of a microscopic
theory develops a dissipative mode 
if the imaginary part of a self-energy is nonzero 
finite in the infrared limit;
 \begin{eqnarray}
  c_1 = \lim_{\p\to \0}\ \lim_{p_0\to 0} \left(\frac{\partial }{\partial p_0}
 \Im{\rm m}\, \Sigma_{\rm R}(p_0,\p) \right) \neq 0, \, \infty \, .
  \label{condition}
 \end{eqnarray}
We propose to check the finiteness of this quantity as a criterion 
for the existence of a dissipative mode. Later we will show that it is indeed 
nonzero finite
at the NLO in $1/N$ expansion in the 2PI formalism.

\begin{figure}
  {\includegraphics[height=2.7cm]{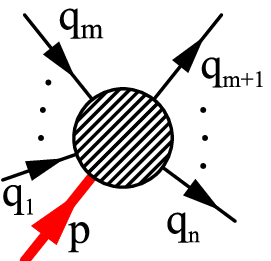}}
\hskip 2cm
  {\includegraphics[height=2.7cm]{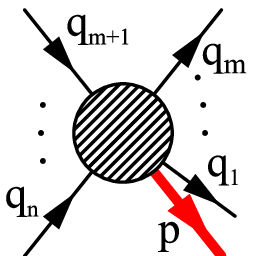}}
\caption{Scattering processes corresponding to cut diagrams in which $n$
internal lines are cut. $m$ runs from $0$ to $n$.}
\label{cut-diagram}
\end{figure}

According to the Cutkosky rules generalized at finite temperature 
\cite{PhysRevD.28.2007, kobes1985discontinuities, kobes1986discontinuities, 
PhysRevD.47.4586, le2000thermal},
one can relate $\Im {\rm m} \Sigma_{\rm R}$ 
to the sum of cut diagrams.
The cut diagrams, in which $n$ internal lines are cut, can be interpreted as 
the scattering processes shown in Fig.~\ref{cut-diagram} 
where an incoming (outgoing) particle with $p^\mu$ 
disappears (emerges)
through the scattering with particles in the heat bath. 
More precisely, shown in the left diagram of Fig.~\ref{cut-diagram} 
is the process where 
a particle with momentum $p$ scatters off $m$ particles 
with momenta $q_1,\cdots,q_m$ in the heat bath, and turns 
to $n-m$ particles with momenta $q_{m+1},\cdots,q_n$ 
in the final state. The right diagram shows the reverse process.

Consider the soft limit of our interest $p\to 0$, where
the energy-momentum conservation 
between the initial and final states, 
$q_1+\cdots+q_m=q_{m+1}+\cdots+q_n$, 
must be fulfilled.
Clearly, the process in which  
there are no incoming particles
from the heat bath (i.e., the initial total energy-momentum is zero)
does not contribute to Eq.~(\ref{condition}) 
because the condition for the final total energy-momentum 
$q_1+\cdots+q_n=0$ can never be satisfied for 
massive particles in the heat bath.
From this simple observation one concludes that 
$\Im {\rm m}\Sigma_{\rm R}(p_0\to 0,\p\to 0)=0$ at $T=0$, 
that is, there is no infrared dissipation in the vacuum. 
At $T\neq 0$, however, the imaginary part of the self-energy 
$\Im {\rm m}\Sigma_{\rm R}(p_0\to 0,\p\to 0)$ 
receives, in general, contributions from the processes 
where 
the incoming particle scatters off particles 
in the heat bath and 
contributions from their reverse processes.

A more careful inspection reveals that
the imaginary part of the self-energy 
$\Im {\rm m}\Sigma_{\rm R}(p_0,\p)$ can be written
in a rather intuitive form:
\begin{eqnarray}&&
\Im {\rm m}\Sigma_{\rm R}(p_0,\p)
\sim
\sum_n
\sum_{m=0}^n \int dq_1\cdots dq_n \, 
\rho^{(m)}(p; q_1,\cdots,q_m) \rho^{(n-m)}(q_{m+1},\cdots,q_n) 
\nonumber\\
&&\hspace{3cm} \times \Big\{{\cal N}[f_{q_1},\cdots, f_{q_n}]
 - \bar{\cal N}[f_{q_1},\cdots, f_{q_n}] \Big\} 
\; ,
\label{rho-rho-n}
\end{eqnarray}
where $\rho^{(m)}$ and $\rho^{(n-m)}$ are, respectively, 
spectral functions for incoming and outgoing thermal particles, 
and ${\cal N}[\{f_i\}]$ is the statistical weight 
of particles participating 
in this process ($\bar{\cal N}[\{f_i\}]$ is for the reverse process). 
The 
weight ${\cal N}[\{f_i\}]$ will be expressed as a product
of thermal distribution functions $f_p=({\rm e}^{\beta|p_0|}-1)^{-1}$. 
The spectral functions specify kinematical windows for scattering states.
Therefore, the quantity $\Im {\rm m}\Sigma_{\rm R}(p_0,\p)$ 
is non-vanishing when the kinematical windows for the incoming and outgoing 
particles have overlap with each other. This is a very useful point of view 
when we evaluate $\Im {\rm m}\Sigma_{\rm R}(p_0,\p)$ in a concrete model.
We will indeed see below that 
$\Im {\rm m}\Sigma_{\rm R}(p_0,\p)$ in the $O(N)$ scalar model 
can be written in this form, and we can 
assess if it is zero or nonzero by checking the kinematical windows.

Note that we will also treat ``composite" fields such as 
$\varphi^2$ since the Schwinger-Dyson equation relates two-point functions 
to higher-point functions, which could have two (or more) fields at 
the same space-time point. Similar consideration as discussed above 
should hold for those ``composite" fields.

\section{Relaxation constant at NLO in $1/N$ expansion}

In this paper, 
as a typical example of relativistic quantum field theories, 
we employ an $O(N)$-symmetric $\varphi^4$ model in $(d+1)$-dimensional
space-time with the action  
 \begin{eqnarray}
  S[\varphi]=\int dtd^dx 
  \left[ \frac{1}{2}\partial_\mu \varphi_a(x) \partial^\mu \varphi_a(x) 
        -\frac{m_0^2}{2}\varphi_a(x)\varphi_a(x) 
        -\frac{\lambda_0}{4!N}(\varphi_a(x)\varphi_a(x))^2\right]\, , 
  \label{micro-action}  
 \end{eqnarray}
where $\varphi_a$ ($a=1,\cdots,N$) is an $N$-component 
real scalar field, $m_0^2$ and $\lambda_0$ are the bare mass and
coupling constant, 
and we assume $d\le 3$ so that the action is renormalizable. 
The partition function in the imaginary time formalism is given by 
 \begin{eqnarray}
  Z[J,K] &=& \int\limits_{\varphi(0,\x)=\varphi(-i\beta,\x)} {\cal D}\varphi
              {\rm exp} 
               \left[
               -S[\varphi] 
               +\int dtd^dx\, J_a(x)\varphi_a(x) 
               +\int dtd^dx dt'd^dy \, \varphi_a(x)K_{ab}(x,y)\varphi_b(y) 
               \right] ,
\notag \\
 \end{eqnarray} 
where $\beta=1/T$, and we have introduced 
one- and two-point source fields, $J$ and $K$, respectively. 
From this partition function $Z[J,K]$, one can introduce the 2PI 
effective action via a double Legendre transform in $J$ and $K$, while
the ordinary 1PI effective action is defined as the single Legendre
transform in $J$ with $K\equiv 0$.
The advantage of the 2PI effective action is that
it reorganizes the perturbative expansion series 
by resuming the higher-order terms self-consistently
into the two-point correlation function.
Details of the formalism of the 2PI effective action are presented in 
Ref.~\cite{Berges:2004yj}.

In the following calculation, we 
employ the $1/N$ expansion, 
which is a typical non-perturbative expansion.
The LO contribution is the so-called tadpole diagram which
only brings in a shift of the mass $m_0^2 \to m^2$. 
Here we restrict ourselves to the symmetric phase: $m^2>0$.
The scattering processes among the particles 
appear at the NLO and in higher order diagrams.
The NLO diagram for the 
self-energy that is relevant for the scattering
processes \footnote{Full diagrams contributing to a 2PI self-energy 
at NLO are given in Ref.~\cite{PhysRevD.66.045008}.} is found to be 
 \begin{eqnarray}
  \Sigma^{\rm scatt.}(i\omega_n,\p)
   &=& 
  \raisebox{-0.4cm}
  {\includegraphics[width=3.9cm]{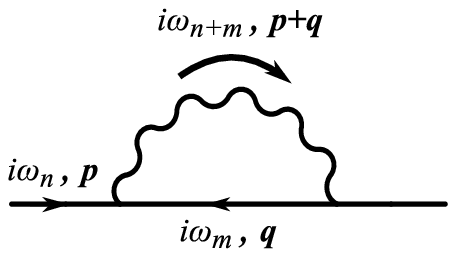}} 
\notag \\
   &=& 
  \raisebox{-0.0cm}
  {\includegraphics[width=2.8cm]{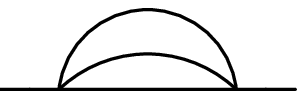}} 
  +
  \raisebox{-0.0cm}
  {\includegraphics[width=3.2cm]{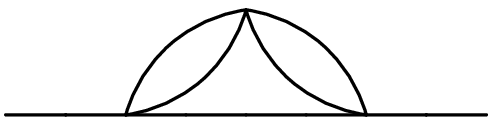}}  
  + \cdots +
  \raisebox{-0.0cm}
  {\includegraphics[width=3.6cm]{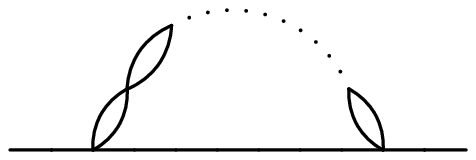}} \ .
  \label{self-energy-0}
 \end{eqnarray}
The straight line and the wiggly line in Eq.~(\ref{self-energy-0}), 
respectively, denote the Matsubara Green function of the elementary field $\varphi$, i.e.
$G \sim \l<\varphi\varphi\r>$, and that of the the composite field $\varphi^2$, i.e.
$D \sim \l<\varphi^2\varphi^2\r>$.
The Matsubara Green function $D(i\omega_n,\p)$ satisfies the following diagrammatic
equation in general:
 \begin{eqnarray}
  D(i\omega_n,\p) 
  \ = \ 
  \raisebox{-0.7cm}
  {\includegraphics[width=4.2cm]{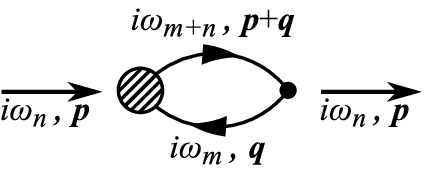}} \ .
  \label{D-graph}
 \end{eqnarray}
The blob in Eq.~(\ref{D-graph}) denotes a vertex function 
$\l<\varphi^2\varphi\varphi\r>$.
This vertex function has been studied with a self-consistent equation 
in Refs.~\cite{Aarts:2004sd, PhysRevD.85.065019}. 
For the purpose of evaluating  the self-energy at NLO, however,
we only need this vertex function at LO  
and then the equation reduces to a geometric series 
of the one-loop bubble diagram $\Pi(i\omega_n, \p)$ 
which leads to the second line of Eq.~(\ref{self-energy-0}): 
 \begin{eqnarray}
  D(i\omega_n,\p)
   &=& \ \ 
  \raisebox{-0.3cm}
  {\includegraphics[width=1.6cm]{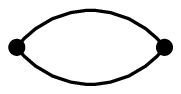}} \ \ 
  + \ \ 
  \raisebox{-0.25cm}
  {\includegraphics[width=5.8cm]{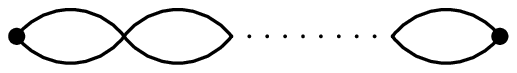}}
  \nonumber \\
   &=& \frac{\Pi(i\omega_n,\p)}
            {1+\frac{\lambda}{6}\Pi(i\omega_n,\p)} 
\; .
  \label{D_R_NLO}
 \end{eqnarray}
 
 At the end of the calculation we perform the analytic continuation from the imaginary
 to the real energy. Then the imaginary-time correlators are replaced by
 their retarded counterparts which are explicitly defined as
 \begin{eqnarray}
  G^{ab}_{\rm R}(p_0,\p) =  G_{\rm R}(p_0,\p) \delta_{ab}
   &=& \int dtd\x \ {\rm e}^{i(p_0t-\p\cdot\x)} \theta(t) \l< [\varphi_a(t,\x) , \varphi_b(0,\0)] \r> , \\
  D_{\rm R}(p_0,\p) 
   &=& \ \int dtd\x \ {\rm e}^{i(p_0t-\p\cdot\x)} \theta(t) \l< [\varphi^2(t,\x) , \varphi^2(0,\0)] \r> .
 \end{eqnarray}
Here $G_{\rm R}^{ab}=G_{\rm R} \, \delta_{ab}$ in the symmetric phase,
and the field indices are suppressed  in $\varphi^2 = \varphi_a\varphi_a$
and 
 \begin{eqnarray}
  D_{\rm R}(p)
   &=& \frac{\Pi_{\rm R}(p)}
            {1+\frac{\lambda}{6}\Pi_{\rm R}(p)} 
\; ,
 \end{eqnarray}
 where
 \begin{eqnarray}
 \Pi_{\rm R}(p)
  &=& \int \frac{dq_0d\q}{(2\pi)^{d+1}} \ n(q_0) \ \rho(q)
\ [G_{\rm R}(p+q)+G_{\rm A}(q-p)]\, .
\label{pi-form}
 \end{eqnarray} 

By using all the information given above, one can find the 
explicit form of the 
imaginary part of the NLO self-energy Eq.~(\ref{self-energy-0})
(see Appendix A for derivation) 
 \begin{eqnarray}
  \Im {\rm m}\Sigma_{\rm R}(p) 
   &=& \frac{\lambda}{6N} \frac{\lambda}{6} 
\int \frac{dq_0d\q}{(2\pi)^{d+1}} \ \rho(q)
  \ \rho_D(p+q) \ [n(q_0)-n(p_0+q_0)] \, ,
  \label{self-energy}
 \end{eqnarray}
where $n(p_0)=({\rm e}^{\beta p_0}-1)^{-1}$,
and  the spectral functions for two-point correlators
are defined as
 \begin{eqnarray}
  \rho(p_0,\p) \equiv 2 \ \Im {\rm m}[G_{\rm R}(p_0,\p)] \ , \qquad \quad
  \rho_D(p_0,\p) \equiv 2 \ \Im {\rm m}[D_{\rm R}(p_0,\p)]
\; ,
  \label{spe-and-im}
 \end{eqnarray} 
Especially, in the LO approximation for $D_R$,
\begin{align}
  \rho_D(p_0,\p)&
 = 
\frac{\  2 \ {\Im} {\rm m} \Pi_{\rm R}(p_0,\p)}
   {\left[ 1+\frac{\lambda}{6} \ {\Re} {\rm e} \Pi_{\rm R}(p)
       \right]^2+ \left[ \frac{\lambda}{6} \ {\Im} {\rm m}
       \Pi_{\rm R}(p_0,\p) \right]^2} 
\; . 
  \label{rho_D} 
\end{align}
Note that the expression Eq.~(\ref{self-energy}) for 
$\Im {\rm m}\Sigma_{\rm R}$ at NLO is valid both in the 
1PI and 2PI formalisms, except that $G$ is the free 
propagator in the 1PI formalism while it is the full 
self-consistent propagator in the 2PI formalism. 
Note also that Eq.~(\ref{self-energy}) has the form as alluded before 
in Eq.~(\ref{rho-rho-n}). Namely, one can judge the finiteness of the 
relaxation constant through the inspection of overlapping kinematical 
windows specified by the spectral functions $\rho$ and $\rho_D$.

In the infrared region, 
$\Im {\rm m}\Sigma_{\rm R}(p)$ 
can be approximated as 
 \begin{eqnarray}
  \Im {\rm m}\Sigma_{\rm R}(p)
   &=& - \frac{\lambda}{6N} \frac{\lambda}{6} 
\int \frac{dq_0d\q}{(2\pi)^{d+1}} \ \rho(q_0,\q) \ \rho_D(q_0,\q) \ p_0\ n'(q_0) +{\cal O}(\p^2, \ p_0^2) .
  \label{self-energy-2}
 \end{eqnarray}
Substituting Eq.~(\ref{self-energy-2}) into Eq.~(\ref{condition}), 
we find the equation which determines the relaxation constant $\Gamma$. 
The aim of the present paper is to show that
$\Gamma$ is nonzero finite.
For notational brevity, 
we define a constant $\gamma$ by
 \begin{eqnarray}
  \frac{1}{\gamma} \equiv - 
\int \frac{dq_0d\q}{(2\pi)^{d+1}} \ \rho(q_0,\q) \ \rho_D(q_0,\q) \ n'(q_0) 
\; ,
  \label{check}
 \end{eqnarray}
which is proportional to $\Gamma$.
In the following subsections, we will examine if 
$\gamma$ is nonzero finite at the NLO of the $1/N$ expansion 
in 1PI and 2PI formalisms.\footnote{
One may wonder if the relaxation constant $\Gamma$ or equivalently 
$\gamma$ defined in Eq.~(\ref{check}) is alternatively defined 
in terms of the current commutators through the Kubo formula 
just like transport coefficients.
Recall however that the Kubo formula evaluates the transport of
conserved densities. On the other hand,
the quantity $\l<\varphi\r>$ here is non-conserved
and can relax locally.
Therefore, the relaxation constant $\Gamma$ is 
naturally given by Eq.~(\ref{condition}) 
or  
Eq.~(\ref{check}). 
}

\subsection{1PI-NLO evaluation}

First, we confirm that a dissipative mode does {\it not} emerge 
in the 1PI two-point function at the NLO. 
In the perturbative expansion in the 1PI formalism,
the propagator $G_{\rm R}$ appearing in the diagrams is
the free propagator obtained at LO with the tadpole effect 
included in the self-energy, whose 
spectral function is
 \begin{eqnarray}
  \rho(p_0,\p) = 2\pi \, {\rm sign}(p_0) \, \delta(p_0^2-\p^2-m^2).
  \label{1PI-spe}
 \end{eqnarray}
The other function $\rho_D(p_0,\p)$ in Eq.~(\ref{check}) 
is the spectral function for a ``composite" $\varphi^2$ field. 
Because the $\lambda \varphi^4$ interaction couples the
$\varphi^2$ field with two $\varphi$ fields 
in either way of $\varphi^2 \varphi\to \varphi$ 
or $\varphi^2 \to \varphi \varphi$ (or their reverse processes),
the spectral function $\rho_D(p_0,\p)$ 
is nonzero in two kinematical regions: 
the space-like region $s=p_0^2-\p^2<0$ and 
two-particle continuum region $s>(2m)^2$. 
The supports of the spectral functions 
$\rho(p_0,\p)$ and $\rho_D(p_0,\p)$ do not overlap,
and therefore the integral (\ref{check})
for $(1/\gamma)_{\rm 1PI}$ at NLO vanishes.
Therefore, we conclude that, in the 1PI-NLO calculation, 
dissipation is not seen in the infrared limit of the $\varphi$ mode.

\begin{figure}[t]
 \begin{center}
  {\includegraphics[width=4.0cm]{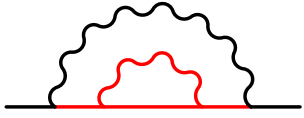}}
 \end{center}
 \caption{
One of the NNLO diagrams in the 1PI calculation. 
The sub-diagram in red 
can be interpreted as a dressing of the propagator $G$,
which gives nonzero spectral strength except for $p_0=0$
on the $p_0$-$\p$ plane.}
 \label{NNLO-diagram}
\end{figure}

One obvious way to proceed is to go beyond the NLO in the 1PI framework.
At the NNLO
the expression for the relaxation constant becomes more involved than
Eq.~(\ref{check}), but it certainly gives a nonzero value
because it is known that the spectral functions $\rho$ and $\rho_D$ 
become nonzero except for $p_0=0$ on the $p_0$-$p$ plane
once the so-called sunset diagram is included in the self-energy 
\cite{PhysRevD.52.3591,blaizot2002quark,PhysRevD.68.076002}. 
The sunset diagram appears as a
sub-diagram in the NNLO-1PI calculation 
(e.g., see Fig.~\ref{NNLO-diagram}, where the wavy line in red includes 
a sunset diagram). 
We will report progress in this approach in a separate paper \cite{xxx}.

In this paper, we take another direction. While keeping the NLO 
approximation, we exploit the 2PI effective action formalism, 
which self-consistently takes into account the processes relevant 
to dissipation in the self-energy of the two-point function $G$.

\subsection{2PI-NLO evaluation}

Let us now confirm that the existence of a dissipative mode is 
accommodated in the 2PI-NLO calculation 
by showing that Eq.~(\ref{check}) is nonzero finite. 
As we have already emphasized several times, 
the 2PI effective action self-consistetnly determines 
two-point functions (and the condensate $\l<\varphi\r>$ 
if $O(N)$ symmetry is broken).
The self-consistent structure is seen in the expression 
(\ref{self-energy}) of $\Im {\rm m}\Sigma_{\rm R}(p_0,\p)$ 
at NLO:
it is given by the spectral functions 
$\rho$ and $\rho_D$, but they are in turn
given as imaginary parts of two-point functions.
Therefore, the precise form of 
$\Im {\rm m}\Sigma_{\rm R}(p_0,\p)$ is found
only after the Schwinger-Dyson-type equation is solved
self-consistently.
We are going to show, however,
from general properties of the spectral functions $\rho$ and $\rho_D$
that $\Im {\rm m}\Sigma_{\rm R}(p_0,\p)$ in 2PI-NLO has a term 
proportional to $p_0$ with a finite coefficient.

Recall that, in the 1PI-NLO calculation, the supports of 
two spectral functions $\rho$ and $\rho_D$ do not have overlap in the 
$(q_0,\q)$ space, giving a vanishing result for Eq.~(\ref{check}), 
but that 1PI-NNLO will give a nonzero finite result thanks to the 
sunset diagram as included in Fig.~\ref{NNLO-diagram}. The same contribution 
is included in the 2PI-NLO diagrams (where we use full propagators) 
and thus Eq.~(\ref{check}) becomes nonzero finite.
Because the spectral functions are odd in the frequency $q_0$
and $n'(x) = n'(-x)= -{\rm e}^x/({\rm e}^x-1)^2$,
Eq.~(\ref{check}) yields
 \begin{eqnarray}
  \frac{1}{\gamma}
   &=& \frac{\Omega_{d}}{(2\pi)^{d+1}} \int_0^\infty dq_0 
       \int_0^\infty q^{d-1}dq \ \rho(q_0,q) \rho_D(q_0,q) \ 
       \frac{{\rm e}^{\beta q_0}}{({\rm e}^{\beta q_0}-1)^2}\, , 
  \label{check2} 
 \end{eqnarray}
where $\Omega_{d}=\frac{2\pi^{d/2}}{\Gamma(d/2)}$ is 
the surface area of the $d$-dimensional sphere\footnote{
The spectral functions are assumed to be rotationally invariant
in equilibrium without external fields. 
}.
Since the spectral functions $\rho$ and $\rho_D$ 
are positive semi-definite for $q_0>0$ in general,
and, as mentioned above, they have nonzero values except for $p_0=0$ 
in the integration domain in the 2PI-NLO calculation
\cite{PhysRevD.52.3591,blaizot2002quark,PhysRevD.68.076002},
we conclude at this point that Eq.~(\ref{check2}) has a nonzero value, 
in constrast to the 1PI-NLO result.


In the following, we verify that Eq.~(\ref{check2}) for $1/\gamma$,  
is not diverging, but finite. To this end, we first notice that the 
spectral functions would not have singularities. For example, 
the delta-function singularity of $\rho$ coming from
the one-particle pole in the vacuum turns into a peak with a finite 
width due to the interactions in the heat bath. Therefore we can 
reasonably assume that the integrand does not have any singularity.

Then, what remains to be checked is the behavior of the 
integrand in the infrared (IR) and ultraviolet (UV) regions. 
For that purpose, we change the variables 
from $(q_0,q)$ to $(Q,\theta)$ defined by 
$q_0=Q\cos\theta,\ q=Q\sin\theta$: 
 \begin{eqnarray}
  \frac{1}{\gamma}
   &=& \frac{\Omega_d}{(2\pi)^{d+1}} \int_0^{\pi/2} d\theta 
        \int_0^\infty QdQ \ (Q\sin \theta)^{d-1} \ 
     \rho(Q,\theta) \rho_D(Q,\theta) \ 
   \frac{{\rm e}^{\beta Q\cos\theta}}{({\rm e}^{\beta Q\cos\theta}-1)^2} . 
  \label{check3'} 
 \end{eqnarray}
In terms of these new variables, the IR and UV limits correspond to 
$Q\to 0$ and $Q\to \infty$, respectively, while there is no 
restriction on the angle $\theta$. Thus, we evaluate the integrand 
in these limiting regions, and then perform the integration over $\theta$ 
to check the finiteness of the integral.

Consider the IR region, $\beta Q\ll 1$, where 
the integrand of Eq.~(\ref{check3'}) is approximated as 
 \begin{eqnarray}
  Q^d \ (\sin \theta)^{d-1} \ 
  \frac{\rho  (Q,\theta)}{Q\cos\theta} \ 
  \frac{\rho_D(Q,\theta)}{Q\cos\theta} . 
  \label{infra-integrand}   
 \end{eqnarray}
In the IR limit $Q\to 0$, spectral functions 
$\rho$ and $\rho_D$ are finite as long as the mass is 
kept finite,\footnote{
However we expect that the spectral function $\rho$ will 
diverge in the IR limit if one considers the massless limit. 
As we will comment later, such a divergent behavior will affect 
the critical dynamics of the two-point functions.
Estimation of this divergence is now under investigation.  }
and thus the quantities $\frac{\rho(Q,\theta)}{Q\cos\theta}$ and 
$\frac{\rho_D(Q,\theta)}{Q\cos\theta}$ are finite 
since $\rho$ and $\rho_D$ are proportional to the frequency 
$q_0=Q \cos\theta$, 
and lastly the $\theta$ integral is also 
finite (for $d>0$). 
Hence, we conclude that no divergence appears from the IR region. 

Next, we do not expect any divergence either
in the UV limit $Q\to \infty$, 
because physically nonzero $1/\gamma$ should be purely 
in-medium effect as is seen in the integrand that contains $n'(q_0)$.
As long as $\theta \neq \pi/2$, the last factor in the integrand
gives an exponential suppression $\sim {\rm e}^{-\beta Q \cos \theta}$
as $Q\to \infty$. 
The remaining part gives at most positive power of $Q$,
as we will see later. 
Therefore, there is no UV divergence for $\theta\neq \pi/2$. 

The only concern is 
the integrand when the angle $\theta$
is close to $\pi/2$, 
namely, in a deep space-like region 
$q_0=Q\cos \theta \to 0,\ q=Q\sin \theta \to \infty$. 
No matter how $Q$ is large, we can 
take 
the quantity $\beta Q\cos\theta$ small enough by 
choosing
$\theta$ very close to $\pi/2$. Then, we find a different estimate 
$\sim 1/(\beta Q \cos \theta)^2 $ for the last factor in the integrand. 
Let us define an angle $\alpha$ which satisfies 
$\beta Q\cos\alpha \ll 1$, 
or equivalently,
${\pi}/{2}-\alpha \ll \frac{1}{\beta Q}$.
We estimate the UV behavior of Eq.~(\ref{check3'}) 
by putting $\alpha$ as the lower limit of the integral: 
 \begin{eqnarray}
  \lefteqn{  \frac{\Omega_d}{(2\pi)^{d+1}} \int_\alpha ^{\pi/2} d\theta \int ^\infty QdQ \ (Q\sin \theta)^{d-1} \ 
     \rho(Q,\theta) \rho_D(Q,\theta) \ \frac{{\rm e}^{\beta Q\cos\theta}}{({\rm e}^{\beta Q\cos\theta}-1)^2}} \nonumber \\
  && \leq  \frac{\Omega_d}{(2\pi)^{d+1}} \int_\alpha ^{\pi/2} d\theta \int ^\infty QdQ \ (Q\sin \theta)^{d-1} \ 
     \rho(Q,\theta) \rho_D(Q,\theta) \ \frac{1}{(\beta Q\cos\theta)^2} .
  \label{check2'} 
 \end{eqnarray}
Since $\rho$ and $\rho_D$ are proportional to $q_0=Q\cos\theta$ 
when $\beta q_0$ is small enough, 
the integrand is regular in the limit $\beta Q\cos\theta \to 0$. 
Let us now introduce two functions as follows:
 \begin{eqnarray}
 h_1(Q)
 \equiv \lim_{\theta \to \pi/2} \frac{\rho(Q,\theta)}{Q\cos\theta} , 
\qquad \quad
h_2(Q) 
\equiv \lim_{\theta \to \pi/2} \frac{\rho_D(Q,\theta)}{Q\cos\theta} . 
  \label{function-1,2}
 \end{eqnarray}
Then, the $\theta$-dependence of the integrand can be neglected, 
and the r.h.s.\ of Eq.~(\ref{check2'}) is evaluated as 
 \begin{eqnarray}
  && \sim  \frac{\Omega_d}{(2\pi)^{d+1}} \ \frac{1}{\beta^2} \int ^\infty 
     \left( \frac{\pi}{2}-\alpha \right) \ QdQ \ Q^{d-1} \ h_1(Q) \ h_2(Q) \, ,
  \label{check2''} 
 \end{eqnarray}
where we have used the approximation $\sin \theta\sim 1$ and 
$\cos\alpha \sim {\pi}/{2}-\alpha $ valid in the vicinity of $\theta =\pi/2$. 
Using the condition ${\pi}/{2}-\alpha \ll \frac{1}{\beta Q}$, 
we can further put an upper limit on the integration 
shown in Eq.~(\ref{check2''}): 
 \begin{eqnarray}
  \lefteqn{ \frac{\Omega_d}{(2\pi)^{d+1}} \ \frac{1}{\beta^2} \int ^\infty 
     \left( \frac{\pi}{2}-\alpha \right) \ dQ \ Q^d \ h_1(Q) \ h_2(Q)} \nonumber \\
  && \ll \frac{\Omega_d}{(2\pi)^{d+1}} \ \frac{1}{\beta^3} \int ^\infty dQ \ Q^{d-1} \ h_1(Q) \ h_2(Q)\, .
  \label{check2'''} 
 \end{eqnarray}
Now the problem has reduced to checking the UV behavior 
of the quantity, 
 \begin{eqnarray}
  Q^{d-1} \ h_1(Q) \ h_2(Q) \, .
  \label{check-r-axis}
 \end{eqnarray}
If this product decays faster than $1/Q$ as $Q\to\infty$, 
the integral is UV-finite.
In order to 
know 
high-momentum behavior of the functions $h_1$ and $h_2$, 
a naive dimensional analysis is sufficient.

Since the mass dimensions of $\rho$ and $\rho_D$ are $-2$ and $0$, 
respectively, those of $h_1$ and $h_2$ in Eq.~(\ref{function-1,2}) 
become $-3$ and $-1$. 
We define two dimensionless functions $\wt h_1$ and $\wt h_2$ 
as follows: 
 \begin{eqnarray}
  h_1=\frac{1}{Q^3} \ \wt h_1 \left( \frac{T}{Q}, \frac{m}{Q},\frac{\lambda}{Q^{3-d}} \right) , \qquad \quad
  h_2=\frac{1}{Q}
\ \wt h_2 \left( \frac{T}{Q}, \frac{m}{Q},\frac{\lambda}{Q^{3-d}} \right)\, , 
\label{naive-counting}
 \end{eqnarray} 
where we have provided the overall mass-dimensions 
by $Q$ and explicitly 
shown the dependence on other dimensionful papameters $T,\ m$ and $\lambda$.
If these parameters enter $\wt h_1$ and $\wt h_2$ 
with positive powers in the deep space-like region,\footnote{
Logarithmic dependence is also possible, but it is weaker than the power 
dependence and thus we ignore such a possibility here.
}
the UV behavior of $h_1$ and $h_2$ is at most determined by the overall 
factor as $h_1\propto 1/Q^3$ and $h_2\propto 1/Q$. Below, we give 
a plausible argument for each parameter to verify that it is indeed 
the case. We will show the finiteness of the functions $h_1$ and $h_2$
(or $\wt h_1$ and $\wt h_2$) in the vanishing limit of parameters $T,\ m$,
and $\lambda$. If the parameters entered the functions with negative 
powers, the functions would diverge at this limit.


First of all, consider the $T$ dependence and the limit $T\to 0$
corresponding to the vacuum (in the absence of a medium). 
In the vacuum, scattering of an incoming particle 
with the heat bath does not occur, and thus only decay of the incoming 
particle (and its reverse process) is physically possible.
Then, 
the spectral functions $\rho$ and $\rho_D$ describe the decay processes 
$\varphi\to \varphi \varphi \varphi$ 
and 
$\varphi^2 \to \varphi \varphi$, respectively. 
Since both have thresholds: $s=q_0^2-\q^2>(3m)^2$ for $\rho$ and $s>(2m)^2$ 
for $\rho_D$, one finds $\rho$ and $\rho_D$ are zero in the space-like 
region $s<0$, and do not diverge in the limit $T\to 0$. 
This means that $\rho$ and $\rho_D$, and so $\wt h_1$ and $\wt h_2$, 
depend on $T$ with positive powers. 

Second, consider the $m$ dependence and the $m\to 0$ limit in 
the space-like region. As long as we stay in a deep space-like region 
where $|\q| \gg m$, the $m$ dependent contributions in two-point 
functions are sub-leading and are given as positive powers of $m/|\q|$ 
(this is evident if one considers a 
free propagator which can be expanded with respect to powers of $m/|\q|$). 
Therefore, in the high momentum region, there is no 
divergence \footnote{This is in contrast with the IR region 
where we expect divergence in the massless limit.}
in spectral functions $\rho$ and $\rho_D$, and thus in $\wt h_1$ and 
$\wt h_2$ when we take the limit $m\to 0$. 

Third  and lastly, consider the $\lambda$ dependence and the limit
$\lambda\to 0$, namely the non-interacting system.  
In this case, $\rho$ is given by the delta-function Eq.~(\ref{1PI-spe}), 
and therefore does not diverge in the space-like region. 
Thus, $\wt h_1$ depends on $\lambda$ with a positive power
in the space-like region. 
The other spectral function $\rho_D$ is given by a single 
bubble diagram (see Eqs.~(\ref{D_R_NLO}), (\ref{spe-and-im})), 
 \begin{eqnarray}
  \rho_D(q)= 2{\Im}{\rm m}\Pi_{\rm R}(q) .
 \end{eqnarray}
Therefore, 
both $\rho$ and $\rho_D$, and thus $h_1$ and $h_2$, 
are finite in the limit $\lambda\to 0$, implying that they are functions
of positive powers of $\lambda/Q$. In the free limit $\lambda\to 0$, we 
can even evaluate the function $h_2$ explicitly.
Note that in this case ($\lambda=0$) the 
two-point function is given by a free propagator: 
 \begin{eqnarray}
  G_{\rm R}(q) = \frac{1}{-(q_0+i\epsilon)^2+\q^2-m^2}, 
 \end{eqnarray} 
where $\epsilon$ is an infinitesimal positive number.  
After analytic calculation, we obtain 
\begin{eqnarray}
  h_2 = \lim_{q_0\to 0} \frac{\rho_D(q)}{q_0} 
     = \frac{1}{4\pi |\q|} \ \frac{1}{{\rm e}^{\sqrt{\q^2+4m^2}/2T}-1}\, .
\end{eqnarray}  
Therefore, the overall mass-dimension is given by 
$1/|\q|\propto Q^{-1}$ which is consistent with the naive dimensional counting
as shown in Eq.~(\ref{naive-counting}),
and $T$ and $m$ dependences appear with positive powers.

Collecting all the results, we conclude that
the high-momentum behavior of $h_1$ and $h_2$ 
are at most $h_1\sim Q^{-3}$ and $h_2\sim Q^{-1}$,
respectively.
Then, 
the dimension of the integrand in Eq.~(\ref{check2'''}) 
becomes $Q^{d-5}$
 and thus the integral is convergent 
for $d<4$ which includes the dimensions of our interest, 
$d\le 3$. 

Hence, in case of the spatial dimension $d\leq 3$, 
Eq.~(\ref{check3'}) becomes nonzero and finite constant. 
Thus, we finally conclude that 
 \begin{eqnarray}
  \left(\frac{1}{\gamma}\right)_{\rm 2PI} \neq 0, \ \infty\, . 
 \end{eqnarray}
This means that a two-point correlation function obtained 
from the self-consistent equation 
in 2PI effective action has a dissipative 
mode in the IR region.

Some comments are in order.
First of all, we emphasize that multi-scattering processes in the 
heat bath are taken into account in the 2PI-NLO but not in the 
1PI-NLO calculation. As a result the dissipative mode, which does 
not exist in the free theory, appears nonperturbatively in the 
2PI-NLO calculation. 
Second, it should be noticed that the dissipative mode emerges
irrespective of whether the microscopic field theory is 
{\it  relativistic} or {\it nonrelativistic}. 
This fact implies the same scaling behavior of the dissipative mode 
for the {\it relativistic} and {\it nonrelativistic} field theory 
at the critical point.
These are in contrast to the propagating modes.
The propagating modes exist in the free theory and 
have different dispersion relations in relativistic and 
nonrelativistic cases. They get modified  
by scatterings in the heat bath to acquire 
damping widths as well as energy shifts. Therefore, 
the scaling behavior of the propagating modes for the {\it relativistic}
and {\it nonrelativistic} field theories is in general different.
Third, we did not attempt to obtain a numerical value of the relaxation
constant, although it is in principle possible 
if one gets the propagators around thermal equilibrium. 
Namely, if one solves the Kadanoff-Baym equation near the equilibrium,
then one can evaluate $1/\gamma$ defined in 
Eq.~(\ref{check}) with the resultant solutions.
We leave this problem for our future work. 
Rather, in the next section, we discuss an important consequence 
deduced from the presence of the diffusion term even without knowing
its coefficient.

\section{Implications for the dynamic exponent $z$}

So far, we have considered the case where the effective mass 
$m^2\equiv m_0^2-\Re {\rm e}\Sigma_{\rm R}(p=0)$ is nonzero. 
Roughly speaking, the limit $m\to 0$ corresponds to the 
critical point where the correlation length diverges
$\xi \sim 1/m \to \infty$. This divergence also affects 
the time evolution of a disturbed system towards equilibrium.
Off the critical point $m\neq 0$, 
small disturbances die away exponentially with the relaxation 
time $\tau$. As the critical point is approached, the relaxation 
time becomes enormously long, 
which is called critical slowing down, and we define 
the dynamic critical exponent $z$ via $\tau\sim \xi^z$ 
\cite{PhysRevLett.19.700,PhysRev.177.952,PhysRevLett.18.891,ferrell1967entropy,ferrell1968fluctuations}.
In this section, we first present a consequence on
the dynamic critical exponent $z$ from the dissipative mode
identified in the previous section, assuming that the relaxation 
constant $\Gamma$ remains nonzero finite 
at the critical point.
Then we remark 
possible modification of such a naive assumption.

In order to identify the dynamic critical exponent $z$, we exploit
the dynamic scaling relation. 
We assume that under a scale transformation
the two-point function transforms anisotropically with respect to 
$p_0$ and $\p$:
 \begin{eqnarray}
  G_{\rm R}(p_0,\p)=b^{2-\eta} \ G_{\rm R}(b^z p_0,b\p) ,
  \label{dynamic-scaling}
 \end{eqnarray} 
where $b$ is a dimensionless scale factor which is taken
bigger than one: $b>1$. The prefactor $b^{2-\eta}$ is fixed by
the static part of the two-point function characterized by an exponent 
$\eta$. 
Indeed, if one substitutes 
$b=\Lambda/|\p|$ into Eq.~(\ref{dynamic-scaling}) and sets $p_0=0$, 
one recovers the static scaling for the correlation 
function $G_{\rm R}^{\rm static}(\p)\sim (\Lambda/|\p|)^{2-\eta}$,
where an arbitrary dimensionful parameter $\Lambda$ is introduced 
to make the combination $\Lambda/|\p|$ dimensionless.
We take $\Lambda$ of the order of the critical temperature $T_c$. 
Next, we will rewrite Eq.~(\ref{dynamic-scaling}) into more useful form to evaluate $z$. 
Since the mass-dimension of $G_{\rm R}$ is $-2$, 
we can write $G_{\rm R}(p_0,\p)$ as \cite{Hohenberg:1977ym}: 
 \begin{eqnarray}
  G_{\rm R}(p_0,\p)=\left(\frac{\Lambda}{|\p|}\right)^{2-\eta} \cdot G_{\rm R} \left( \frac{\Lambda^z p_0}{|\p|^z} , \Lambda \right) 
   = \Lambda^{-2}\left(\frac{\Lambda}{|\p|}\right)^{2-\eta} \cdot \wt G_{\rm R} \left( \Lambda^{z-1} \frac{ p_0}{|\p|^z} \right) \, .
  \label{spe-scaling}
 \end{eqnarray} 
Thus, we can extract $z$ from the different scaling properties of
$p_0$ and $\p$ in the two-point function.

The self-consistent equation obtained in the 2PI framework is
previously utilized to evaluate the static critical exponents 
$\eta$ \cite{Alford:2004jj}
and $\nu$ \cite{PhysRevD.85.065019} 
(see Ref.~\cite{PhysRevD.15.2897, vanHees:2002bv, Alford:2004jj} for 
the application of the 2PI framework to 2nd-order phase transitions). 
While the imaginary part
of the self-energy $\Im {\rm m}\Sigma_{\rm R}(p)$ vanishes
in the static limit,
the momentum dependence of the real part $\Re {\rm e}\Sigma_{\rm R}(p)$
brings in the scaling form of the two-point function:
 \begin{eqnarray}
  G_{\rm R}^{\rm static}(|\p|;T_c) 
= \Lambda^{-2} \left(\frac{\Lambda}{|\p|}\right)^{2-\eta} . 
 \end{eqnarray}
Of course, this is the same form as identified in the prefactor in 
Eq.~(\ref{dynamic-scaling}). Since this scaling function is simply obtained 
from $\Re {\rm e} \Sigma_{\rm R}(p)$, we can substitute it into the two-point
function (\ref{hydro-prop}) to find the expression at the critical point:
 \begin{eqnarray}
  G_{\rm R}(p;T_c) = \frac{1}{-i\Gamma^{-1}p_0+{\Lambda^\eta}|\p|^{2-\eta}} 
  = \Lambda^{-2}\left(\frac{\Lambda}{|\p|}\right)^{2-\eta}\cdot 
\frac{1}{-i\Gamma^{-1}\Lambda^{-\eta}p_0/|\p|^{2-\eta}+1} \, .
  \label{G-critical}
 \end{eqnarray}
This yields the dynamic critical exponent
 \begin{eqnarray}
  z=2-\eta \, ,
  \label{exponent}
 \end{eqnarray} 
which depends on $N$ and $d$ through $\eta$ 
\cite{PhysRevLett.32.1413, Alford:2004jj, Vasil'ev-text:2004}.

In order to better understand the result (\ref{exponent}), 
let us carefully see how it appears. For that purpose, it is 
instructive to go back to the 1PI-NLO calculation which however 
does not produce a dissipative mode in the IR region. Absence of 
a dissipative mode in 1PI-NLO is deduced from the IR behavior of 
$\Im{\rm m}\Sigma_{\rm R}(p)$, namely in the IR region it starts 
from ${\cal O}(p_0^2,\ \p^2)$ instead of ${\cal O}(p_0)$. Even though
we do not have a dissipative mode, we are able to define a ``dynamic 
exponent" from the dispersion relation in the IR region. However, 
this is an exponent for the ``propagating mode" and must be distinguished 
from the exponent of a dissipative mode which governs very long-time
evolution of the system. Indeed, if one considers a propagator 
in the 1PI-NLO calculation $[G_{\rm R}^{{\rm 1PI-NLO}}(p)]^{-1} \sim 
{-p_0^2+\p^2-\Re {\rm e}\Sigma_{\rm R}^{{\rm 1PI-NLO}}(p)
-i\Im{\rm m}\Sigma_{\rm R}^{{\rm 1PI-NLO}}(p)} $
with $\Im{\rm m}\Sigma_{\rm R}^{{\rm 1PI-NLO}}(p)={\cal O}(p_0^2,\ \p^2)$,
one finds dispersions for ``propagating modes" $p_0\sim \pm |\p|$ 
as the leading contribution with small correction from the self-energy
of order ${\cal O}(1/N)$. Thus, one will find an exponent 
 $z_{\rm prop.}=1+{\cal O}(1/N)$, where we have introduced 
$z_{\rm prop.}$ to remind that it is not the exponent for 
a dissipative mode, but for a propagating mode. If one performs 
the same analysis in 
{\it nonrelativistic} field theories, one will find 
 $z_{\rm prop.}=2+{\cal O}(1/N)$ again for propagating modes whose 
leading dispersions are $p_0\propto \p^2$.
Namely, the leading contribution $z_0$ in the exponent 
$z_{\rm prop.}=z_0+{\cal O}(1/N)$ depends on whether the system 
is relativistic or nonrelativistic, and this is one of the reasons 
why there was confusion in the literature. Now, coming back to the 
2PI-NLO calculation where we do have a dissipative mode, we find that 
the presence of a dissipative term, 
$-i\Gamma^{-1}p_0$, modifies the dispersion, and the leading contribution
to $z$ is given by 2. 
Higher order correction $-\eta$ to $z=2$ is from 
$\Re {\rm e}\Sigma_{\rm R}(p)$ and it is of ${\cal O}(1/N)$. 
Therefore, our result $z=2-\eta$ in Eq.~(\ref{exponent}) looks 
similar to the one for nonrelativistic propagating modes 
$z_{\rm prop.}=2+{\cal O}(1/N)$, but the origin is different.

Let us further proceed one more step. 
Our result (\ref{exponent}) is an immediate consequence of the static 
scaling behavior under the assumption that the relaxation term 
$-ip_0/\Gamma$ remains non-singular \cite{PhysRev.95.249, PhysRev.95.1374}.
However, the relaxation constant $\Gamma$ may depend on $\p$ 
in general.
If one assumes 
$\Gamma\sim \Gamma_0 \, (|\p|/\Lambda)^{c'}$ with  $\Gamma_0$
and $c'$ being constants in the IR region, then one obtains 
 \begin{eqnarray}
  G_{\rm R}(p_0,\p) \sim \frac{1}{-i\Gamma_0^{-1}(\Lambda/|\p|)^{c'}p_0 
+\Lambda^\eta\p^{2-\eta}} .
\label{modified}
 \end{eqnarray}
and accordingly the exponent $z$ is modified to
 \begin{eqnarray}
  z=2-\eta +c'. \label{modified-z}
 \end{eqnarray}
This is also clearly seen in the $t$-dependence of the two point function. 
Equation (\ref{modified}) leads to $G_{\rm R}(t,\p)\sim \exp \{ -t/\tau(\p)\}$
where the relaxation time is given by 
$\tau(\p)\sim 1/(\Gamma(\p) \p^{2-\eta}) \sim 1/\p^{2-\eta+c'}$. 
Thus, in the IR region $\p\to 0$, the relaxation time diverges.
We are now working on the evaluation of the numerical value of $c'$,
and the result will be reported soon elsewhere \cite{xxx}. 

Our result (\ref{modified-z}) with modified $\Gamma$ is 
consistent with ``model A" in the classification for 
nonrelativistic systems based on the mode-coupling 
theory \cite{Hohenberg:1977ym}. 
Model A contains only (multi-component) 
non-conserved fields without conserved fields, and the dynamic critical 
exponent is given by $z=2+c\eta$, 
where $c$ is 
a numerical constant of order one 
\cite{Hohenberg:1977ym, PhysRevLett.29.1548}. 
It seems apparently  to be the case 
since we are only looking at the fundamental field $\varphi$. 
But it is actually nontrivial because 
in the 2PI framework the energy-momentum tensor is conserved and the
coupling between the non-conserved fields $\varphi$ and conserved
fields is relevant to the dynamic universality classification.
However, without knowing the explicit numerical value for $c'$, 
it is difficult at present to judge to which model our result is classified. 

There is an explicit numerical simulation for a particular case 
with $N=1$ and $d=2$ \cite{Berges:2009jz}. From numerical evaluation of the 
dynamic critical exponent $z\sim \frac{2-\eta}{0.87}$, it is claimed that 
it belongs to the dynamic universality class, model C. 
However, since we have used the $1/N$ expansion, which is not valid 
for $N=1$, 
we cannot directly compare ours to the numerical simulation result,
which 
is intriguing , though.

\section{Summary}

Long-time behavior of a system recovering equilibrium should be characterized
by a dissipative hydrodynamic mode. 
Because of the non-linear mixing among the modes, 
we expect that even the two-point function $G$ 
of the order parameter field $\varphi$ carries 
the information of this long-time dissipation.

We have shown, however, that the NLO calculation of $G$ 
in the 1PI $1/N$ expansion only gives real and imaginary corrections 
to the propagating mode of $\delta \varphi$,
but it does not generate the dissipative mode in $G$,
especially that the imaginary part of the self-energy, 
${\Im}{\rm m}\Sigma_{\rm R}(p_0,\p)/p_0$, vanishes in the IR limit,
$p_0, \p \to 0$.
At the critical point, then we find that the mode spectrum scales as
$p_0 \propto p^{z_{\rm prop.}}$ with 
$z_{\rm prop.}=z_0+{\cal O}(1/N)$, where $z_0=1$ and 2
for relativistic and nonrelativistic cases, respectively. 
Since there is no dissipative mode in the IR limit, 
this exponent is just associated with the propagating mode, 
whose dispersion depends on whether the system is relativistic 
or nonrelativistic.

In the 2PI framework, on the other hand, 
we resumed the multiple scattering  effects 
in the self-consistent propagator $G$ although working 
at NLO in $1/N$ expansion. 
Once the self-consistency is imposed on $G$,
the spectral function $\rho(p_0,\p)$ 
will become nonzero except for $p_0=0$ 
even at the NLO in the $1/N$ expansion.
(Recall the sunset diagram for example which is included at this order.)
Then we have argued that the dissipative mode inevitably
emerges as an IR pole of the propagator $G$ by showing that
the imaginary part of the self-energy ${\Im}{\rm m}\Sigma_{\rm R}(p_0,\p)/p_0$ is 
generally nonzero finite in the IR limit.

At the critical point, this dissipative mode implies that 
we will have $z \sim 2-\eta$ even in relativistic scalar theories,
which is in contrast to the 1PI-NLO result. 
For more precise evaluation of the exponent $z$, one may need
to compute the transport coefficient utilizing the 4PI framework.
We leave this as our future work.
Furthermore, it is not clear whether the dissipative mode we found
is the same as the dissipation of the hydrodynamic mode such as heat.
To this end, we apparently need
more detailed and systematic evaluation of
the coupled two-point functions of $\varphi$ 
and the energy-momentum densities, etc.

\section*{Acknowledgments}
The authors are very grateful to J\"urgen Berges 
for useful discussions and correspondence 
on critical dynamics
of $O(N)$ scalar theory. 
YS and OM would like to thank Yoshimasa Hidaka for 
discussions on non-equilibrium phenomena in general.
HF's work was partially supported by Grant-in-Aid for Scientific Research (C) No. 24540255.

\appendix

\section{Derivation of Eq.~(\ref{self-energy})}

In this Appendix, we derive Eq.~(\ref{self-energy}) 
in spatial $d$-dimensions using imaginary time formalism. 
As we have graphically shown in Eq.~(\ref{self-energy-0}), 
the (scattering part of the) self-energy at NLO is explicitly written as 
 \begin{eqnarray}
  \Sigma(i\omega_n,\p) = \frac{\lambda}{3N} \ \frac{\lambda}{6} \ T
   \sum_m \int \frac{d\q}{(2\pi)^d} 
  \ G(i\omega_m,\q) \ D(i\omega_{n+m},\p+\q)\, ,
  \label{matsubara-sigma} 
 \end{eqnarray}
where $i\omega_n$ is the Matsubara frequency, 
$i\omega_n=2\pi i n/\beta \ (n\in \mathbb{Z})$. 
In the 1PI formalism, $G$ and $D$ are {\it free} correlation 
functions, and summation over the Matsubara frequency is easily 
transformed into a contour integral on a complex plane of an 
analytically continued variable $\zeta$. On the other hand, 
in the 2PI formalism, $G(i\omega_n,\q)$ and $D(i\omega_n,\q)$ 
are {\it full} correlation functions and, 
when analytically continued, both $G(\zeta,\q)$ and $D(\zeta,\q)$ 
have cuts along the real axis $\Im {\rm m}\, \zeta =0$ except for 
the origin $\zeta=0$. Namely, $G(\zeta,\q)$ and $D(\zeta,\q)$ have 
discontinuities along the real axis, but one can define $G(0,\q)$
and $D(0,\q)$ without ambiguity. Thus, we have to be careful when 
we transform the Matsubara summation into a contour integral over 
the complex variable $\zeta$ \cite{holstein1964theory,PhysRevD.66.045005}.
This was done in Ref.~\cite{Aarts:2004sd} when the spatial dimension 
$d$ is three. Here, we will follow Ref.~\cite{Aarts:2004sd}, but work
in $d$ dimensions.


In the complex plane of $\zeta$, one can recover the Matsubara 
summation by picking up all the residues of the function, 
$n(\zeta)=({\rm e}^{\beta\zeta}-1)^{-1}$. 
Therefore, Eq.~(\ref{matsubara-sigma}) becomes 
 \begin{eqnarray}
  \Sigma(i\omega_n,\p) = \frac{\lambda}{3N} \ \frac{\lambda}{6} \ 
   \int_{\cal C} \frac{d \zeta}{2\pi i} 
   \int \frac{d\q}{(2\pi)^d} 
   \ n(\zeta) \ G(\zeta,\q) \ D(i\omega_n +\zeta,\p+\q) ,
  \label{contour-1}
 \end{eqnarray}
where the integral-contour ${\cal C}$ is a collection of circles 
going 
around counterclockwise each pole of $n(\zeta)$: 
$\zeta=2\pi im/\beta \ (m\in \mathbb{Z})$. 
We have to be careful when we pick up the poles at $\zeta=0$
and $\zeta=-i\omega_n$ because there are cuts originating from the same 
points. The easiest way to avoid this is to infinitesimally 
shift the poles so that one can safely pick up the residues. This simple
prescription indeed works and the result does not depend on the way 
how to shift the poles.

\begin{figure}[t]
 \begin{center}
  \includegraphics[width=60mm]{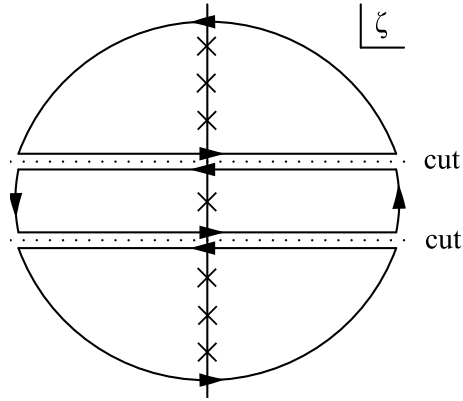}
 \end{center}
 \caption{
The 
 complex plane of $\zeta$. 
Dotted lines along ${\Im}{\rm m} \ \zeta=0$ and 
${\Im}{\rm m} \ \zeta=-\omega_n$ represent cuts of 
$G(\zeta)$ and $D(i\omega_n+\zeta)$, respectively. 
The crosses denote poles of $n(\zeta)$, all of 
which are on the imaginary axis. 
The contour ${\cal C}$ can be modified to three closed paths 
 ${\cal C}'$ (solid lines). 
}
 \label{complex-plain}
\end{figure}

Then, we modify the contour ${\cal C}$ (collection of small 
circles around the poles of $n(\zeta)$) paying attention
\footnote{
$G(\zeta)$ has no poles in ${\Im}{\rm m} \ \zeta \neq 0$ 
\cite{baym1961determination, blaizot2002quark, le2000thermal}, 
and the same argument can be applied to $D(\zeta)$. 
Thus, we need not beware singularities of $G(\zeta)$ and $D(\zeta)$ 
except for their discontinuities along the real axis. 
} 
to the presence of cuts along $\zeta=0$ and $\zeta=-i\omega_n$, 
and obtain a new contour ${\cal C'}$ as shown in Fig.~\ref{complex-plain}.

In this integral, contributions from 
infinite distances become zero. 
Therefore, we pick up the integrals along two discontinuities. 
Namely, we consider four paths: 
from $\zeta=-\infty \pm i\epsilon$ to $\zeta=+\infty \pm i\epsilon$, 
and from $\zeta=-\infty -i\omega_n \pm i\epsilon$ to $\zeta=+\infty -i\omega_n \pm i\epsilon$ 
($\epsilon$ is an infinitesimal positive number). 
Thus, we find that Eq.~(\ref{contour-1}) becomes 
 \begin{eqnarray}
  \Sigma(i\omega_n,\p) 
   &=& \frac{\lambda}{3N} \ \frac{\lambda}{6} \ \int_{-\infty}^\infty \frac{dq_0}{2\pi i} \int \frac{d\q}{(2\pi)^d} \ n(q_0) \nonumber \\
   && \times \left\{ [G(q_0+i\epsilon,\q) -G(q_0-i\epsilon,\q)]\ D(i\omega_n +q_0,\p+\q) \right. \nonumber \\
   && \quad \left. + G(q_0-i\omega_n,\q) \ [D(q_0+i\epsilon,\p+\q) -D(q_0-i\epsilon,\p+\q)] \right\} . 
  \label{contour-2}  
 \end{eqnarray}
In order to obtain the retarded self-energy, 
we perform analytic continuation $i\omega_n \to p_0+i\epsilon$. 
 \begin{eqnarray}
  \Sigma_{\rm R}(p_0,\p) 
   &=& \frac{\lambda}{3N} \ \frac{\lambda}{6} \ \int \frac{dq_0d\q}{(2\pi)^{d+1}} \ n(q_0) \nonumber \\
   && \times \left\{ \rho(q_0,\q) \ D_{\rm R}(p_0+q_0,\p+\q) 
     +G_{\rm A}(q_0-p_0,\q) \ \rho_D(q_0,\p+\q) \right\} ,
  \label{contour-3}  
 \end{eqnarray}
where we have used the following properties 
 \begin{eqnarray}
  \rho(q_0,\q)= -i \ [G_{\rm R}(q_0,\q)-G_{\rm A}(q_0,\q)] , \qquad \ 
  \rho_D(q_0,\q)= -i \ [D_{\rm R}(q_0,\q)-D_{\rm A}(q_0,\q)] \ . 
 \end{eqnarray}
Now, we take the imaginary part of Eq.~(\ref{contour-3}). 
Since the spectral functions are imaginary parts of correlation functions, 
 \begin{eqnarray}
  \rho(p) = 2\ {\Im}{\rm m} \ G_{\rm R}(p) = -2\ {\Im}{\rm m}\ G_{\rm A}(p)\, ,
 \ \qquad 
  \rho_D(p) = 2\ {\Im}{\rm m} \ D_{\rm R}(p) = -2\ {\Im}{\rm m} \ D_{\rm A}(p) \ , 
 \end{eqnarray}
we finally obtain Eq.~(\ref{self-energy}): 
 \begin{eqnarray}
  \Im {\rm m}\Sigma_{\rm R}(p_0,\p) 
   &=& \frac{\lambda}{6 N} \ \frac{\lambda}{6} \ \int \frac{dq_0d\q}{(2\pi)^{d+1}} \ n(q_0) \nonumber \\
   && \times \left\{ \rho(q_0,\q) \ \rho_D(p_0+q_0,\p+\q) 
     -\rho(q_0-p_0,\q) \ \rho_D(q_0,\p+\q) \right\} \nonumber \\
    &=& 
   \frac{\lambda}{6 N} \ \frac{\lambda}{6} \ \int \frac{dq_0d\q}{(2\pi)^{d+1}} \ \rho(q) \ \rho_D(p+q) \ [n(q_0)-n(p_0+q_0)] .
  \label{contour-4} 
 \end{eqnarray}
In the second line of Eq.~(\ref{contour-4}), 
we have changed the integration variables of the second term as 
$q_0\to q_0+p_0$ 
and $\q \to \p+\q$.



 \end{document}